\documentclass[prl,twocolumn, letterpaper]{revtex4}
\usepackage{amsmath}
\usepackage{amssymb}
\usepackage{graphicx}
\usepackage{hyperref}

\begin{document}

\title{Wave Dark Matter Non-minimally Coupled to Gravity}
\author{Lingyuan Ji}
\email{lingyuan.ji@jhu.edu}
\affiliation{Department of Physics and Astronomy, Johns Hopkins University\\ 3400 North Charles Street, Baltimore, MD 21218, United States}

\begin{abstract}
    We consider a model where a light scalar field (with mass $\lesssim 30\, {\rm eV}$), conjectured to be dark matter, has a non-minimal coupling to gravity. In the non-relativistic limit, this new coupling introduces a self-interaction term in the scalar-field equation of motion, and modifies the source term for the gravitational field. Moreover, in the small-coupling limit justified by the observed dark-matter density, the system further reduces to the Gross-Pitaevskii-Poisson equations, which remarkably also arise from a self-gravitating and self-interacting Bose-Einstein condensate system. We derive predictions of our model on linear and non-linear structure formation by exploiting this unexpected connection.
\end{abstract}

\maketitle

With decades of compelling evidence about its existence, the identity of dark matter remains perhaps as the central question in cosmology \cite{Bertone:2004pz}. Proposals for dark matter range from an ultra-light axion with mass $\sim 10^{-22}\,{\rm eV}$ \cite{Hui:2016ltb, Hu:2000ke} to primordial black holes with mass $\sim 10 M_\odot$. A distinction occurs at around mass $\sim 30\,{\rm eV}$ \cite{Hui:2021tkt}, below which the mean separation of dark matter particles is smaller than their de~Broglie wavelength, rendering their wave-like behavior important.

Motivated by various particle physics considerations, there are many flavors of these wave dark matter models, most of which consist of a scalar field that is minimally coupled to gravity. However, the scalar field could, or some \cite{Faraoni:2000wk} would argue in general has to, be non-minimally coupled to gravity. It has been shown that a non-minimal coupling will naturally arise as quantum corrections to a minimally coupled classical theory \cite{Birrell:1982ix}. Moreover, a non-minimal coupling term is crucial to the renormalizability of a scalar-field theory in curved space-time \cite{Callan:1970ze, Freedman:1974ze, Freedman:1974gs}. This coupling will change the dynamical behavior of dark matter and can potentially have observable effects in structure formation.

Here, we consider a non-minimal coupling of the form $\phi^2 R$, where $\phi$ is the scalar field and $R$ is the Ricci scalar. We first write down the general theory and derive the equations of motion in the non-relativistic limit. We point out the difference between this theory and the minimally coupled theory of wave dark matter. We then discuss the small-coupling limit valid in most practical cases, while making connections to self-gravitating and self-interacting Bose-Einstein condensate \cite{Chavanis:2011gm, Chavanis:2011uv, Chavanis:2011zi, Chavanis:2011zm, Visinelli:2017ooc, Colpi:1986ye, Widdicombe:2018oeo}. Following those connections, we present predictions of this model on linear and non-linear structure formation. We end with several concluding remarks.

We consider the theory $S=S_{\rm EH} + S_\phi$, where
\begin{equation}\label{eqn:action-einstein-hilbert}
    S_{\rm EH} = \int d^4x \sqrt{-g}\, \frac{R}{16\pi G}
\end{equation}
is the familiar Einstein-Hilbert action, and 
\begin{equation}\label{eqn:action-phi}
	S_{\phi} = \int d^4x \sqrt{-g}\, \left[-\frac12 (\nabla^\mu\phi) (\nabla_\mu\phi) - V(\phi) - \frac12 \xi R \phi^2 \right]
\end{equation}
is the scalar-field action with a non-minimal coupling to gravity. Here, $g \equiv \det(g_{\mu\nu})$ is the determinant of the metric tensor $g_{\mu\nu}$; $V(\phi)$ is the potential of the scalar field; and $\xi$ is a dimensionless coupling constant. The equation of motion for the scalar field is determined by $\delta S/\delta \phi = 0$, which in this case gives
\begin{equation}\label{eqn:eom-phi}
	\Box \phi - \xi R \phi - V'(\phi) = 0.
\end{equation}
Here, $\Box\equiv g^{\mu\nu}\nabla_\mu \nabla_\nu$ is the d'Alembertian with $\nabla_\mu$ being the covariant derivative, and $V'(\phi)\equiv dV(\phi)/d\phi$. In the rest of this paper, we assume $\phi=0$ is a local minimum of the potential, and the excursion of $\phi$ is small enough so that we can write $V(\phi) = m^2\phi^2/2$. We define the energy-momentum tensor of the scalar field as $T^\phi_{\mu\nu} \equiv -(2/\sqrt{-g}) \delta S_\phi/\delta g^{\mu\nu} $ and obtain
\begin{multline}\label{eqn:energy-momentum-phi}
	T^\phi_{\mu\nu} = (\nabla_\mu \phi)(\nabla_{\nu} \phi) - g_{\mu\nu}\left[\frac12 (\nabla^\alpha \phi)(\nabla_\alpha \phi) + V(\phi)\right] \\+ \xi \left[G_{\mu\nu} \phi^2 + g_{\mu\nu} \Box(\phi^2) - \nabla_\mu \nabla_\nu(\phi^2) \right].
\end{multline}
It is worth noting that the variation of the last term in Eq.~(\ref{eqn:action-phi}) is subject to the Leibniz rule, and thus more complex than the variation of the Ricci scalar $R$ in Eq.~(\ref{eqn:action-einstein-hilbert}). This explains the three terms proportional to $\xi$ in Eq.~(\ref{eqn:energy-momentum-phi}) \cite{Kodama:1985bj}. The Einstein equation $\delta S/\delta g^{\mu\nu}=0$ can then be written as
\begin{equation}\label{eqn:einstein-equation}
    G_{\mu\nu} = 8\pi G T^\phi_{\mu\nu}.
\end{equation}
Here, $G_{\mu\nu}\equiv R_{\mu\nu} - R g_{\mu\nu}/2$ is the Einstein tensor. Note that, due to the non-minimal coupling, the right-hand side of Eq.~(\ref{eqn:einstein-equation}) also contains the geometric quantity $G_{\mu\nu}$.

We consider the weak-gravity limit in the Newtonian gauge with only the scalar metric perturbation $\Psi$, where the line element is
\begin{equation}
    ds^2 = - (1+2\Psi)dt^2 + (1-2\Psi)\delta_{ij}dx^i dx^j.
\end{equation}
Note that we ignore the cosmic expansion for simplicity. The effect of cosmic expansion can be simply restored by considering ``Newtonian cosmology'' \cite{Chavanis:2011uv}, provided that the pressure is negligible in comparison with the energy density. We also neglect any anisotropic stress so that the other metric perturbation variable $\Phi$ is equal to $\Psi$, and the gravity sector is described by only one variable $\Psi$. This will be consistent with the non-relativistic limit we are about to take later in this work. From now on, we assume $|\Psi|\ll 1$ and only work to leading order in $\Psi$.

This space-time geometry implies some formulas that will be useful later. For a scalar quantity $f = f(\vec x, t)$, the d'Alembertian is
\begin{equation}\label{eqn:box-explicit}
    \Box f= (1+2\Psi)\nabla^2f - (1-2\Psi)\partial_t^2f + 4 (\partial_t \Psi) (\partial_t f);
\end{equation}
the $(0,0)$ component of the Hessian is
\begin{equation}\label{eqn:hessian-00-explicit}
    \nabla^0\nabla_0f = \nabla \Psi \cdot \nabla f + (\partial_t \Psi) (\partial_t f) - (1-2\Psi)\partial_t^2 f;
\end{equation}
the Ricci scalar is
\begin{equation}\label{eqn:ricci-explicit}
    R = 2\nabla^2 \Psi - 6\partial_t^2\Psi;
\end{equation}
and the $(0,0)$ component of the Einstein tensor $G^{\mu}_{\ \nu}$ is
\begin{equation}\label{eqn:einstein-00-explicit}
    G^0_{\ 0} = -2\nabla^2 \Psi.
\end{equation}
Note that here $\nabla$ without the subscript is the flat spatial gradient operator, and should not be confused with $\nabla_\mu$.

In the non-relativistic limit we will discuss, only two equations are important. The first equation is the scalar-field equation of motion. By inserting Eq.~(\ref{eqn:box-explicit}) (with $f=\phi$) and Eq.~(\ref{eqn:ricci-explicit}) into Eq.~(\ref{eqn:eom-phi}), we have
\begin{multline}\label{eqn:eom-phi-explicit}
    (1+2\Psi)\nabla^2\phi - (1-2\Psi)\partial_t^2 \phi - m^2\phi \\ + 4 (\partial_t \Psi) (\partial_t \phi) - \xi(2\nabla^2 \Psi - 6\partial_t^2\Psi)\phi= 0.
\end{multline}
The second equation is the $(0,0)$ component of the Einstein equation $G^0_{\ 0} = 8\pi G T^0_{\ 0}$. By inserting Eqs.~(\ref{eqn:box-explicit}) and (\ref{eqn:hessian-00-explicit}) (with $f=\phi^2$) plus Eq.~(\ref{eqn:einstein-00-explicit}) into Eq.~(\ref{eqn:einstein-equation}), we have
\begin{equation}\label{eqn:einstein-explicit}
    \nabla^2\Psi = 4\pi G \rho_\phi,
\end{equation}
where the energy density $\rho_\phi \equiv -T^0_{\ 0}$ of the scalar field is
\begin{multline}
    \rho_\phi = \frac12 (1-2\Psi)(\partial_t \phi)^2 + \frac12 (1+2\Psi)(\nabla\phi)^2 + \frac12 m^2 \phi^2\\
    + \xi \big[ 2(\nabla^2 \Psi)\phi^2 + (\nabla \Psi)(\nabla \phi^2) \\
    - (1+2\Psi)(\nabla^2 \phi^2) - 3(\partial_t \Psi)(\partial_t \phi^2) \big].
\end{multline}
Note that due to the non-minimal coupling, $\rho_\phi$ itself now contains $\nabla^2\Psi$. This implies that the metric perturbation $\Psi$ is not sourced by $\rho_\phi$, but a slightly more complicated term. Later, in the non-relativistic limit, we will show exactly how the source term is modified.

We now work out the non-relativistic limit of the equation of motion, Eq.~(\ref{eqn:eom-phi-explicit}), and the Einstein equation, Eq.~(\ref{eqn:einstein-explicit}), by factoring out the fast-varying oscillation $e^{-imt}$ in $\phi$ as
\begin{equation}\label{eqn:wkb}
    \phi = \frac{1}{\sqrt{2m}}\left(\psi e^{-imt} + \psi^* e^{+imt}\right).
\end{equation}
The newly defined complex scalar $\psi$ is then slowly varying (i.e.\ $\partial_t \ll m$ when acting on everything other than $\phi$). The non-relativistic limit also implies that the gradient of $\phi$ (and $\psi$) is small (i.e.\ $\nabla \ll m$ when acting on $\phi$ or $\psi$). Now, we proceed by working in the appropriate orders of $\Psi$, $\partial_t/m$, and $\nabla/m$. We shall also average out the fast-varying contribution to the energy density $\rho_\phi$. Since these standard procedures have been detailed for a minimally coupled scalar field in Ref.~\cite{Marsh:2015xka}, we only explain in detail how we approximate terms proportional to $\xi$. In Eq.~(\ref{eqn:eom-phi-explicit}), we only keep $\nabla^2 \Psi$ but neglect $\partial_t^2 \Psi$. This is justified by the smallness of the spatial part of the Einstein equation (i.e.\ $|G^i_{\ j}| \ll |G^0_{\ 0}|$) in the non-relativistic limit. In Eq.~(\ref{eqn:einstein-explicit}), we only keep $(\nabla^2 \Psi)\phi^2$. The terms $(\nabla \Psi)(\nabla \phi^2)$ and $(1+2\Psi)(\nabla^2 \phi^2)$ are neglected due to the smallness of $\nabla\phi$; the term $(\partial_t \Psi)(\partial_t \phi^2)$ is approximately $(\partial_t \Psi)(\partial_t |\psi|^2/m)$ after averaging out the fast-varying contribution, and is then neglected due to the smallness of $\partial_t \psi$.

Doing so, Eqs.~(\ref{eqn:eom-phi-explicit}) and (\ref{eqn:einstein-explicit}) then become the coupled differential equations
\begin{align}
    i\partial_t \psi &= -\frac{\nabla^2}{2m}\psi + m \Psi \psi + \xi \mathcal V(\psi) \psi, \\
    \nabla^2 \Psi &= m \mathcal V(\psi),
\end{align}
where we define the effective potential,
\begin{equation}
    \mathcal V(\psi) \equiv \frac{4\pi G m |\psi|^2}{m-8\pi \xi G |\psi|^2}.
\end{equation}
It is obvious that without non-minimal coupling ($\xi=0$), the system further reduces to the Schr\"odinger-Poisson equations. In the presence of non-minimal coupling, a self-interaction term is added to the Schr\"odinger equation, and the source term of the Poisson equation is modified. As will become clear later, we are practically always in the small-non-minimal-coupling limit $\xi \ll m/(8\pi G|\psi|^2)$. This implies $\mathcal V(\psi) \approx 4\pi G|\psi|^2$, and the system reduces to the Gross-Pitaevskii-Poisson equations
\begin{align}
    i\partial_t \psi &= -\frac{\nabla^2}{2m}\psi + m \Psi \psi + 4\pi \xi G |\psi|^2 \psi; \label{eqn:gross-pitaevskii}\\
    \nabla^2 \Psi &= 4\pi G m|\psi|^2. \label{eqn:poisson}
\end{align}
From this familiar form, we can then interpret $m|\psi|^2$ as the dark-matter density $\rho$ (not to be confused with $\rho_\phi$). So, retrospectively, the small coupling limit is valid when
\begin{equation} \label{eqn:small-xi-limit}
    \xi \ll \frac{m^2}{8\pi G\rho} \simeq 8 \times 10^{15}\left(\frac{m}{10^{-22}\, {\rm eV}}\right)^2\left(\frac{1\, {\rm GeV/cm^3}}{\rho}\right),
\end{equation}
which is easily satisfied if $\xi$ is not unnaturally large. We proceed with the small-coupling limit from now on.

Interestingly, the Gross-Pitaevskii-Poisson equations also arise in the description of self-gravitating and self-interacting Bose-Einstein condensate \cite{Chavanis:2011gm, Chavanis:2011uv, Chavanis:2011zi, Chavanis:2011zm, Visinelli:2017ooc}. Although the equations are equivalent, the physical picture is completely different. In the Bose-Einstein condensate scenario, the self-interaction term $\propto |\psi|^2\psi$ is the direct result of the explicitly introduced contact interaction between the bosons, while in our case it arises from the scalar field's non-minimal coupling to gravity. This unexpected correspondence allows us to benefit from the results of previous work. For instance, we provide the translation from Ref.~\cite{Chavanis:2011zi} to our paper,
\begin{equation}\label{eqn:translation}
    \hbar\to 1,\quad N\to 1, \quad g\to \frac{4\pi G}{m^2} \xi.
\end{equation}
The last one can be substituted by $a \to G m \xi$. On the left-hand side, $\hbar$ is the reduced Planck constant; $N$ is the number of particles in the condensate; and $g$ is the contact-interaction strength, with the s-wave scattering length $a$ being an equivalent representation. We emphasize that the duality established here is only in the appropriate limits of both theories (non-relativistic and small-coupling limit for the theory presented in this work, and non-relativistic limit for the Bose-Einstein condensate). While the duality likely does not exist in the full theory, relativistic studies of the self-interacting Bose-Einstein condensate (for instance, Refs.~\cite{Colpi:1986ye, Widdicombe:2018oeo}), once reduced to the non-relativistic limit, should re-obtain this duality.

The Gross-Pitaevskii-Poisson equations, Eqs.~(\ref{eqn:gross-pitaevskii}) and (\ref{eqn:poisson}), have a fluid description. We define the fluid variables
\begin{equation}
    \rho \equiv m|\psi|^2, \quad \text{and} \quad \vec v \equiv \frac{1}{m}\nabla \arg\psi.
\end{equation}
It can be shown that Eqs.~(\ref{eqn:gross-pitaevskii}) and (\ref{eqn:poisson}) are equivalent to the following fluid equations
\begin{align}
    \partial_t \rho + \nabla \cdot (\rho \vec v) &= 0, \label{eqn:fluid-density}\\
    \partial_t \vec v + (\vec v \cdot \nabla)\vec v &= - \nabla \Psi - \frac{\nabla P_\xi}{\rho} - \frac{\nabla\cdot P_Q}{\rho}, \label{eqn:fluid-velocity}\\
    \nabla^2 \Psi &= 4\pi G \rho. \label{eqn:fluid-poisson}
\end{align}
These equations can also be obtained by translating Eqs.~(10), (14), and (7) in Ref.~\cite{Chavanis:2011zi} using our Eq.~(\ref{eqn:translation}). Here we define the quantum pressure tensor and the pressure caused by non-minimal coupling as
\begin{equation}
    P_{Q, ij} \equiv - \frac{\rho}{4m^2}\partial_i \partial_j \ln \rho, \quad \text{and} \quad P_\xi \equiv \frac{2\pi \xi G}{m^2}\rho^2,
\end{equation}
respectively. While we have a fully classical system here, we still choose to follow the widely accepted name of $P_Q$ as the quantum pressure. Note that the non-minimal coupling amounts to an isotropic pressure term, whereas the quantum pressure is anisotropic. We refer to $P_\xi$ as the ``$\xi$-pressure'' from now on.

The Jeans scale $k_J$ of this theory is crucial to linear structure formation. Intuitively, gravity pulls the dark matter together and makes structure grow, while the quantum pressure and the $\xi$-pressure resist this effect. Quantitatively, we examine the divergence of the right-hand side of Eq.~(\ref{eqn:fluid-velocity}) by setting $\rho = \bar\rho (1+\delta)$ and working to leading order in $\delta$, which gives (in the original sequence of terms)
\begin{equation}\label{eqn:fluid-velocity-rhs}
    - 4\pi G \bar\rho \delta - \frac{4\pi\xi G\bar\rho}{m^2} \nabla^2 \delta + \frac{1}{4m^2} \nabla^4 \delta.
\end{equation}
Here, in the first term, we use the Poisson equation, Eq.~(\ref{eqn:fluid-poisson}), with only the perturbation $\bar\rho\delta$, without the background $\bar\rho$, as the source (i.e.\ the ``Jeans swindle'' \cite{Falco:2012ud}). Substituting $\nabla$ with $i\vec k$, we see that a given $k$-mode of density perturbation $\delta$ will grow normally (as cold dark matter) at low $k$ when gravity wins, but this growth is suppressed at high $k$ when the quantum pressure and the $\xi$-pressure win. The transition, namely the Jeans scale $k_J$, follows from solving the resulting quadratic equation in $k^2$. Doing so, we find,
\begin{equation} \label{eqn:jeans-scale}
    k_J = (16\pi G \bar\rho)^{\frac14} m^{\frac12} \left( \sqrt{1+\frac{4\pi G\bar\rho}{m^2}\xi^2} - \sqrt{\frac{4\pi G\bar\rho}{m^2}\xi^2} \right)^{\frac12}.
\end{equation}
Note that this formula can also be obtained by translating Eq.~(138) in Ref.~\cite{Chavanis:2011zi} using our Eq.~(\ref{eqn:translation}). The dimensionless quantity
\begin{multline} \label{eqn:jeans-deviation}
    \frac{4\pi G \bar \rho}{m^2} \xi^2 \simeq 8 \times 10^{-25} \left(\frac{\xi}{0.1}\right)^2 \left(\frac{10^{-22}\, {\rm eV}}{m}\right)^2 \\ \times \left(\frac{\bar\rho}{1.3 \times 10^{-6}\, {\rm GeV}/{\rm cm}^3}\right)
\end{multline}
determines $k_J$'s deviation from the Jeans scale $k_{J0} = (16\pi G \bar\rho)^{\frac14} m^{\frac12}$ of the minimally coupled theory, and is likely a small number if $\xi$ is not unnaturally large. This can be understood, in a different way, by considering the scales $k_\xi$ and $k_Q$ at which the quantum pressure and the $\xi$-pressure, respectively, become comparable to gravity. By comparing the second and the third terms to the first term in Eq.~(\ref{eqn:fluid-velocity-rhs}), we find
\begin{equation}
    k_\xi = \frac{m}{\sqrt\xi}, \quad \text{and} \quad k_Q = (16\pi G \bar\rho)^{\frac14} m^{\frac12} = k_{J0}.
\end{equation}
Taking the same anchored values as in Eq.~(\ref{eqn:jeans-deviation}), we see that $k_\xi^{-1} \sim 0.02\,{\rm pc}$ is far smaller than $k_Q^{-1} \sim 0.014\,{\rm Mpc}$, so the new $\xi$-pressure is not important in modifying the Jeans scale $k_{J0}$ ($= k_Q$) of the minimally coupled theory. But, we need to point out that, in principle, it is still possible for a large enough $\xi$, without violating the small-coupling limit in Eq.~(\ref{eqn:small-xi-limit}), to make Eq.~(\ref{eqn:jeans-deviation}), hence the deviation of $k_J$ from $k_{J0}$, large.

\begin{figure}
    \centering
    \includegraphics[width=\linewidth]{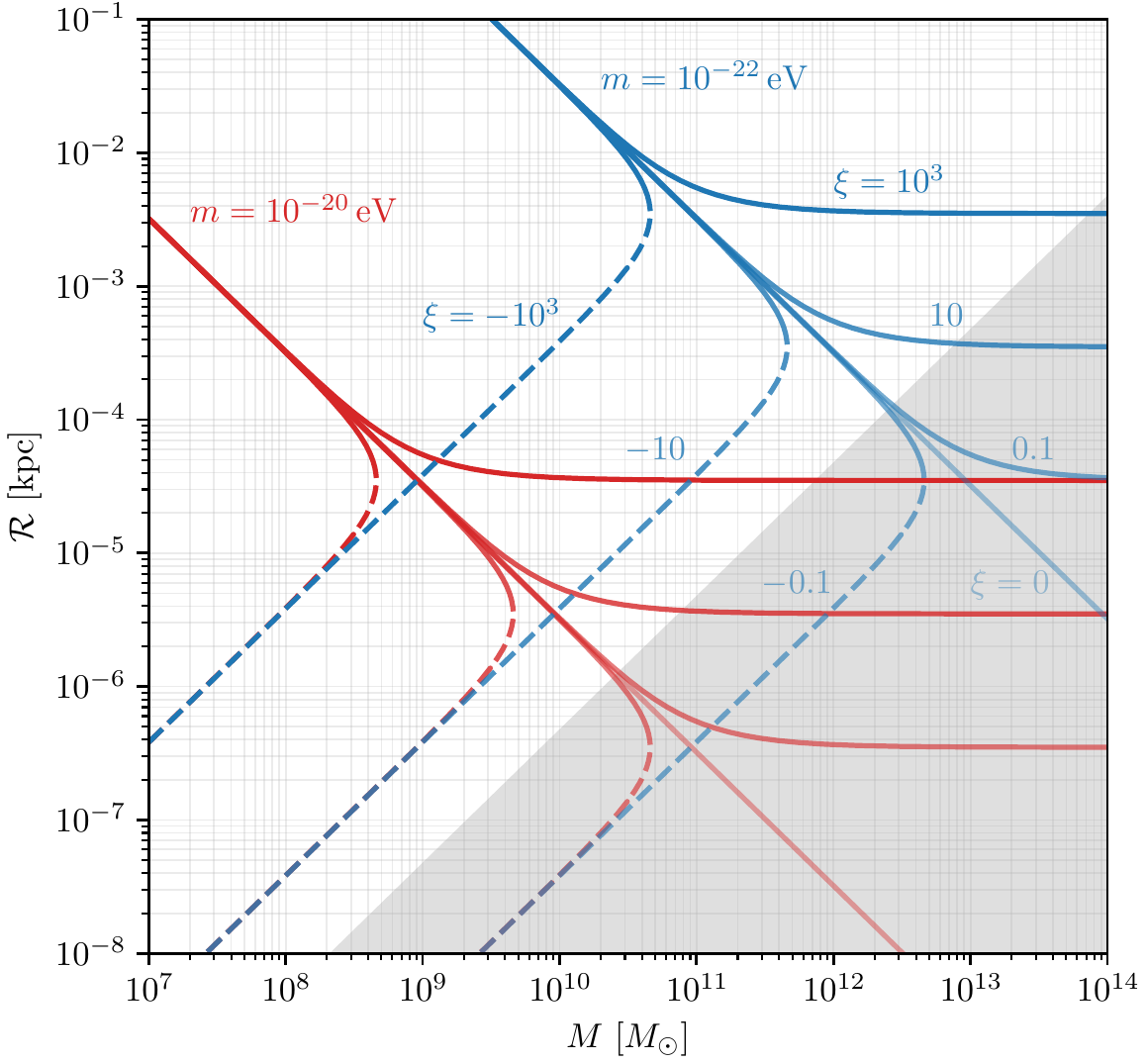}
    \caption{Mass-radius relationship, Eq.~(\ref{eqn:mass-radius}), of the soliton for different dark-matter mass $m$ and non-minimal coupling $\xi$. The blue (red) curves show mass $m=10^{-22}\, {\rm eV}$ ($10^{-20}\, {\rm eV}$). Among those curves, $\xi$ takes the values $\{+10^3,+10,+0.1,0,-0.1,-10,-10^3\}$, with $\xi=0$ giving the minimal coupling limit. For negative values of $\xi$, the solid lines indicate the choice of plus sign in Eq.~(\ref{eqn:mass-radius}), while the dashed lines indicate minus sign. We note that the minus-sign branch when $\xi<0$ is dynamically unstable \cite{Chavanis:2011zi}. The shaded region indicates $\mathcal R < GM$, where the relativistic effect becomes important, and the soliton might collapse into a black hole \cite{Nazari:2020fmk, Muia:2019coe, Helfer:2016ljl}. We expect our non-relativistic treatment to become increasingly insufficient approaching this region.}
    \label{fig:mass-radius}
\end{figure}

However, the $\xi$-pressure will almost always be important at small scales, implying noticeable change in non-linear structure formation. In wave dark matter models, numerical simulations show that the dark-matter halo will host an enormous soliton --- as massive as $10^8 M_\odot$ for a $10^{10} M_\odot$ halo with $m\sim 10^{-22}\,{\rm eV}$ \cite{Schive:2014dra, Schive:2014hza}. The profile of a soliton for a given mass can be obtained by solving the hydrostatic version of Eqs.~(\ref{eqn:fluid-density}), (\ref{eqn:fluid-velocity}), and (\ref{eqn:fluid-poisson}) \cite{Chavanis:2011zm}. It is also possible to obtain approximate analytical results by assuming a Gaussian density profile $\rho(r) = M(\pi \mathcal R^2)^{-3/2} \exp(-r^2/\mathcal R^2)$ of the soliton, and minimize the energy functional with respect to the characteristic soliton size $\mathcal R$ for a given soliton mass $M$ \cite{Chavanis:2011zi}. The two approaches are found to yield similar results \cite{Chavanis:2011zm}. Here, we translate the Bose-Einstein-condensate result, Eq.~(92) in Ref.~\cite{Chavanis:2011zi}, using our Eq.~(\ref{eqn:translation}), giving the mass-radius relation
\begin{equation}\label{eqn:mass-radius}
    \mathcal R = \frac{(9\pi/2)^{1/2}}{GMm^2}\left[\frac12 \pm \frac12 \sqrt{1+\frac{8\xi}{3\pi}(GMm)^2}\right].
\end{equation}
The ``$\pm$'' only takes the plus sign for $\xi>0$, but takes the plus and the minus signs for $\xi<0$. We emphasize that the minus-sign branch when $\xi<0$ is dynamically unstable \cite{Chavanis:2011zi} (i.e.\ a local maximum of the energy functional). Representative evaluations of this formula are given in Fig.~\ref{fig:mass-radius}. We see that for the dynamically stable branches, a positive $\xi$ increases the radii of large-mass solitons, whereas a negative $\xi$ decreases them. We also see that, for a negative $\xi$, there is a maximum mass and a minimum radius for a stable soliton. \emph{In addition} to this limit, the soliton cannot become too compact (i.e.\ $\mathcal R \sim GM$), at which point our non-relativistic treatment will become insufficient, and the soliton might collapse into a black hole \cite{Nazari:2020fmk, Muia:2019coe, Helfer:2016ljl}.

Now, the mass-radius relationship, Eq.~(\ref{eqn:mass-radius}), can be used to interpret observational results. We only present two preliminary examples here, and leave a full-fledged study to future work. First, a recent dynamical analysis of the Galactic center suggests a $10^9\, M_\odot$ solitonic core with a size of  $0.1\, {\rm kpc}$ \cite{DeMartino:2018zkx}. This would only be compatible, in the minimally coupled theory, with $m \simeq 10^{-22}\, {\rm eV}$. When allowing non-minimal coupling, interpretations with different $m$ emerge when $|\xi|$ becomes large ($\gtrsim 10^4$, but still within the small-coupling limit). Second, the minimally coupled theory predicts the mass-radius scaling $\mathcal R \sim M^{-1}$, in tension with the observed constant core surface density for various low-mass galaxies \cite{Burkert:2020laq} (i.e.\ $M/\mathcal R^2 \sim {\rm const.}$, implying $\mathcal R \sim M^{1/2}$). Although the non-minimal coupling cannot fully resolve the problem, it will alleviate it by providing a $\mathcal R \sim M^0$ scaling via the plateau part of Eq.~(\ref{eqn:mass-radius}) at large soliton mass.

Before closing, we identify some similarities between the theory presented here, in the small-coupling limit, and other models of wave dark matter. A minimally coupled ($\xi=0$) theory with an explicit self-interaction $\lambda \phi^4 \in V(\phi)$ in Eq.~(\ref{eqn:action-phi}) will also yield similar Gross-Pitaevskii-Poisson equations, Eqs.~(\ref{eqn:gross-pitaevskii}) and (\ref{eqn:poisson}) \cite{Ferreira:2020fam}. This can be understood by rewriting the non-minimal coupling $\phi^2 R$ in the non-relativistic limit --- $\phi^2 R \sim \phi^2 \nabla^2 \Psi \sim \phi^4 $. Here, the first relation is implied by Eq.~(\ref{eqn:ricci-explicit}), and the second by the leading contribution from time averaging the Poisson equation, Eq.~(\ref{eqn:poisson}). This also means that our theory can be described by the non-relativistic effective field theory for scalar dark matter, detailed recently in Ref.~\cite{Salehian:2021khb}.

In this work, we discuss a theory of wave dark matter that has a non-minimal coupling $\phi^2 R$ to gravity. We derive the equations of motion for this theory in the non-relativistic and small-coupling limit and present an equivalent fluid description with the Gross-Pitaevskii-Poisson equations. From that, we also point out a connection between this theory and previous research on self-gravitating and self-interacting Bose-Einstein condensate. We proceed to discuss some phenomenology of linear and non-linear structure formation. Future work may explore the next-to-leading-order effect in $\xi$; the cosmological matter power spectrum incorporating the full expansion history of the Universe; the production process in the early Universe; numerical simulations of halo formation, etc. It should also be interesting to see the consequences of other forms of non-minimal coupling.

The author would like to thank M.~Kamionkowski, D.~Grin, and T.~Helfer for helpful discussions and comments on an earlier draft. This work was supported in part by NSF Grant No.~1818899.

\end{document}